\documentclass{eptcs}
\providecommand{\event}{Axel Legay (Ed.): International Workshop on Verification of Infinite-State Systems (INFINITY)} 
\providecommand{\volume}{13}   
\providecommand{\anno}{2009}   
\providecommand{\firstpage}{1} 
\providecommand{\eid}{1}       
\usepackage{breakurl}        
\usepackage{macro}
\usepackage{picinpar}
\usepackage{epsfig}

\title{Automated Predicate Abstraction for \\ Real-Time Models}

\author{Bahareh Badban\thanks{Corresponding author} \qquad\qquad Stefan Leue
\institute{Department of Computer and Information Science, University of
  Konstanz, Germany}
\email{\quad Bahareh.Badban@uni-konstanz.de \quad\qquad Stefan.Leue@uni-konstanz.de}
\and
Jan-Georg Smaus
\institute{Institut f\"ur Informatik, Universit{\"a}t Freiburg, Germany}
\email{Smaus@informatik.uni-freiburg.de}
}
\def\titlerunning{Automated Predicate Abstraction for Real-Time Models}
\def\authorrunning{B. Badban \& S. Leue \& J.G. Smaus}
\begin{document}
\maketitle
\thispagestyle{empty}
\setcounter{page}{\firstpage}

\paragraph{Introduction}
Model checking has been widely successful in validating and debugging hardware
designs and communication protocols. However, state-space explosion is an
intrinsic problem which limits the applicability of model checking tools. To
overcome this limitation software model checkers have suggested different
approaches, among which abstraction methods have been highly esteemed.  modern
techniques.  Among others, predicate abstraction is a prominent technique which
has been widely used in modern model checking.  This technique has been shown
to enhance the effectiveness of the reachability computation technique in
infinite-state systems.
In this technique an infinite-state system is represented abstractly by a
finite-state system, where states of the abstract model correspond to the
truth valuations of a chosen set of atomic predicates.
Predicate abstraction was first introduced in~\cite{Graf97} as a method for
automatically determining invariant properties of infinite-state systems.
This technique involves abstracting a concrete transition system using a set
of formulas called {\em predicates} which usually denote some state properties
of the concrete system.

The practical applicability of predicate abstraction is impeded by two problems.
First, predicates need to be provided manually~\cite{Lahiri06b,Das02}. 
This means that the selection of appropriate abstraction predicates is based
on a user-driven trial-and-error process. The high degree of user intervention
also stands in the way of a seamless integration into practical software 
development processes. Second, very often the abstraction is too coarse in order
to allow relevant system properties to be verified. This calls for abstraction 
refinement~\cite{Colon98}, often following a counterexample guided abstraction refinement
scheme~\cite{clarke00guided,Ball02}.
Real time models are one example of systems with a large state space as time
adds much complexity to the system. In this event, recently there have been
increasing number of research to provide a means for the abstraction of such
models.  It is the objective of this paper to provide support for an automated
predicate abstraction technique for concurrent dense real-time models
according to the timed automaton model of~\cite{alur94}.  We propose a method
to generate an efficient set of predicates than a manual, ad-hoc process would
be able to provide.  We use the results from our recent work~\cite{BaLeSm09}
to analyze the behavior of the system under verification to discover its local
state invariants and to remove transitions that can never be traversed.
We then describe a method to compute a predicate abstraction based on these
state invariants. We use information regarding the control state labels as
well as the newly computed invariants in the considered control states when
determining the abstraction predicates.  We have developed a prototype tool
that implements the invariant determination.  Work is under way to also
implement the computation of
a predicate abstraction based on our proposed method. We plan to embed 
our approach into a comprehensive abstraction and refinement methodology
for timed automata.

\paragraph{Related Work.}
An interactive method for
predicate abstraction of real-time systems where a set of predicates called
{\em basis} is provided by the user is presented in~\cite{Colon98}. 
The manual choice of the abstraction basis depends on the user's understanding
of the system.  The work presented in~\cite{Moller02,Sorea:FTRTFT04} proposes an abstraction method
which is based on identifying a set of predicates that is fine enough to
distinguish between any two clock regions and which creates a strongly preserving
abstraction of the system. 
The basis predicates are discovered by spurious paths obtained
through model-checking of the system. Also, in this approach the choice of
the original set of predicates relies on the user's understanding of the system, 
as well as on the counterexample generation experiments.
To the best of our knowledge, at the time of writing, there
has been no research done on {\em automatically} 
generating invariants (predicates) for dense real time models,
which will be the central contribution of our paper.

In the functional setting the CEGAR methodology based on the seminal 
paper~\cite{clarke00guided}
has been rather influential in the development of hard- and software 
verification methodologies, e.g.,~\cite{Ball02}.
Abstraction predicate discovery based on the analysis of 
spurious counterexamples is at the heart of the work 
in~\cite{Das02}.
The approaches presented in~\cite{Jhala05,millan06} and  
in~\cite{Henzinger04} use interpolation to detect feasibility of an 
abstract trace. \cite{millan03} introduces a
proof-based automatic predicate abstraction.

\section{Preliminary Definitions and our Previous Results}
\label{sec.ta}

\paragraph{Timed Automata.}
To have this article self-contained we need to briefly explain some of the
results in~\cite{BaLeSm09}.  A {\em timed automaton}
~\cite{bluebook,alur94}.consists of a finite state automaton together with a
finite set of clock variables, simply called {\em clocks}, and a finite set of
integer variables.  In the notation we distinguish clock and integer variables
only where necessary.  Clocks are non-negative real valued variables which all
increase at the same speed, while integers change only when there is an
explicit assignment.  Initially, all clocks are set to $0$.  A clock may be
reset, but afterwards it immediately starts running again.  The finite state
automaton describes the system {\em control} states of the system, which are
referred to as \emph{locations}, as well as its transitions between locations.
A {\em state} or configuration of the system has the form $\< l,u\>$ where $l$
is the current control location and $u$ is a valuation function which assigns
to each its current value.  For $d{\in} {\mathbb{R^+}}$, we denote by $u+d$ a
valuation that assigns to each clock $x$ the value $u(x)+d$, i.e., it
increases the value of all clocks by $d$, while the integer variables remain
unchanged.
$\G(X)$ denotes the set of (clock or integer) {\em constraints} $g$ for a set
  $X$ of clock variables. Each $g$ is of the form $g:=x\leq t~|~t\leq
  x~|~\lnot g~|~ g_1\land g_2,$ where $x\in X$, and $t$, called {\em term}, is
  either a variable in $X$ or a linear integer expression, which is an
  expression of the form $c+\sum_{i=1}^nc_i\cdot x_i$ where the $x_i$ are
  integer variables and $c$ and $c_i$ are integer constants.\footnote{The
    restriction to integers does not constitute a loss of generality
    \cite[Section 4.1]{alur94}.}  We usually write $s<t$ for $\lnot\; t\leq
  s$. By $var(g)$ we denote the set of all clock variables appearing in $g$.
%
Formally, a timed automaton $\A$ is a tuple $\< L,l_0,\Sigma,X,\I,E \>$ where 
\begin{itemize}
\item $L$ is a finite set of {\em (control) locations}. $l_0\in L$ is the initial location.
\item $\Sigma$ is a finite set of labels, called {\em events} or {\em channels}.  
\item $X$ is a finite set of variables. 
\item 
$\I:L\mapsto \G(X)$ assigns to each location in $L$ some 
constraint in $\G(X)$.   
\item $E\subset L\times \Sigma\times 2^{X}\times \G(X)\times L$ represents
  {\em discrete} transitions.
\end{itemize}

The constraint associated with each location $l\in L$ is called its {\em
  invariant}, denoted $\I(l)$. We later refer to these invariants as the {\em
  original} invariants. Time can pass in a control location $l$ only as long
as $\I(l)$ remains $\tr$, i.e. $\I(l)$ must hold whenever the current location
is $l$.
%
The semantics of a nondeterministic timed automaton $\A$ is defined by a {\em transition
  system} $\SA$. 
States or configurations of $\SA$ are
pairs $\<l,u\>$, where $l\in L$ is a control location of $\A$ and $u$ is a
valuation over $X$ which satisfies $\I(l)$, i.e. $u\models \I(l)$. 
$\<l_0,u\>$ is an {\em initial} state of $\SA$ if $l_0$ is the initial location.

\noindent
{\em Transitions.} 
For each transition system the system state changes by:
\begin{itemize}

\item {\em Delay transitions}, denoted by $d$, which allow time $d{\in} {\mathbb{R^+}}$ to 
elapse. The value of all clocks is increased by $d$ leading to the
transition $\< l,u\> \stackrel{d}{\mapsto} \< l,u+d\>$.
\footnote{Recall that the integer variables remain unchanged.} 
This transition can take place only when the invariant of location $l$ is
satisfied along the transition, i.e.\ $\forall d'\leq d: u+d'\models \I(l)$.

\item {\em Discrete transitions}, denoted by $\tau$, which enable a transition. A transition
  $\tau$ is {\em enabled} when the current clock valuation satisfies
  $\G_{\tau}$.  When $\tau$ is executed, all variables, except those which are
  reset, remain unchanged. This results in the transition $\tau:= \< l,u\>
  \stackrel{a,g,r}{\mapsto} \< l',u'\>$ where $a$ is an event, $g$ is a guard
  and $r$ is a reset.
\end{itemize}
An {\em execution} of a system is a possibly infinite sequence of
states 
$\< l,u\>$ where each pair of two consecutive states corresponds to
either a discrete or a delay transition.


\paragraph{Creating New Invariants by $\cipm$.}
Here, we explain briefly the $\cipm$ algorithm from ~\cite{BaLeSm09}. This
algorithm strengthens the given original invariants in each control location
by analysing the incoming discrete transitions to that specific control
location; It also reduces the size of the model by pruning away those
transitions which can never be traversed.
The input of the $\cipm$ algorithm is a timed automaton $\A$, the output is
$\A$'s pruned version together with a set of new invariants for $\A$.

A discrete transition $\tau: \< l,u\>\mapsto\< l',u'\> $ is called {\em idle}
if it can never be enabled.  Amongst other reasons, a transition can be idle
when the constraint over the transition is unsatisfiable, or when the
valuation function obtained from the transition does not satisfy the invariant
of the target location, which means that $u'\not \models \I(l')$.  For
instance, if $\tau$ is the discrete transition $\<l,u\>\stackrel{x\leq
  y}{\mapsto}\<l',u'\>$ where $x> y+3$ is an invariant in location $l$, then
this transition is idle since the constraint $x\leq y$ is never satisfied.

At each control location $l_i$, $\cipm$ first collects the set $\I(l_i)$ of
all the original invariants, and then accumulates all its incoming transitions
in $\int(l_i,\A)$.  The idle transitions within these sets are identified and
are deleted from the model.

For each non-idle $\tau$ in $\int(l_i,\A)$ the algorithm next computes all
propagated constraints into $l_i$.  Since $l_i$ may also have some original
invariant, the new invariant, i.e. $\Ia(l_i)$, is the conjunction of the
original invariant and all of the previously computed imposed constraints on
$l_i$.  Computing $\Ia(l_i)$ may render some of the outgoing transitions of
$l_i$ idle.  Therefore, the algorithm next checks all outgoing transitions of
$l_i$ for idleness again.  It then removes all transitions detected as being
idle.
Two timed automata $\A$ and $\A_1$ are {\em equivalent}, denoted $\A\eqa\A_1$,
if they differ only in some idle transitions.
\begin{theorem}
\label{theo.inv}
$\cipm$ always terminates. It also satisfies the following properties:
\begin{itemize} 
\item if $\cipm(\A_1)=(\A,\Ia)$ then $\A\eqa\A_1$.
\item If $\cipm(\A_1)=(\A,\Ia)$, then $u\models\Ia(l)$, for each reachable
  state $\<l,u\>$ in $\SA_1$.  In other words, $\Ia(l)$ is invariant in $l$.
\end{itemize}
\end{theorem}

\paragraph{Networks of Timed Automata.}

$\cipm$ can also be used to treat networks of timed automata in which several
parallel automata synchronize with one another via synchronous message passing.
Transitions associated with emitting or receiving a message of type $a$ are labeled
with $!a$ or $?a$, respectively. 
The intuitive semantics of a synchronous message passing is such that the message sending and the 
message receiving primitives are blocking and executed in a rendez-vous manner.

Formally, the semantics of this kind of synchronization is defined as follows.
Let $A=\<\bar{L},\bar{l^0} ,\Sigma,X,\I,E\>$ be a parallel composition of $n$
timed automata $\A_1,\dots,\A_n$, denoted by $\A= \A_1\|\dots\|\A_n$, where
$\A_i:=\< L_i,l^0_i ,\Sigma_i,X_i,\I_i,E_i \>$ for each $1\leq i\leq n$ and
for each two non-equal $i$ and $j$ $X_i\cap X_j=\emptyset$.  For $A$ we have
$X=\bigcup_{1\leq i\leq n} X_i$, $\Sigma=\bigcup_{1\leq i\leq n} \Sigma_i$,
and $\I(\bar{l})= \bigwedge_{1\leq i\leq n} \I(l_i)$ for $\bar{l}= (l_1,\dots,
l_n)$. The initial location is denoted by $\bar{l^0}= (l^0_1,\dots, l^0_n)$.
A state of the network is a configuration $\<\bar{l}, u\>$ where
 $\<l_i,u_i\>$ is a configuration in $\A_i$ and $u(x)=u_i(x)$ for each $x\in X_i$ and $1\leq i\leq n$.
$\bar{l}[l_i/l'_i]$ denotes the replacement of $l_i$ by $l'_i$ in $\bar{l}$,
which is $\bar{l}[l_i/l'_i]=(l_1,\dots, l_{i-1},l'_i, l_{i+1},\dots,l_n)$.
Delay transition in this systems is defined as before. Other transitions are:
\begin{itemize}
%
\item {\em Discrete transitions:} If 
$\< l_i,u_i\> \stackrel{a,g,r}{\mapsto}\< l'_i,u'_i\>$ then $\tau:=\<\bar{l},
u\>\stackrel{a,g,r}{\mapsto}\<\bar{l}[l_i/l'_i], u'\>$ is a discrete
transition in the network model if $u'(x){=}u'_i(x)$ for $x{\in} X_i$ and
$u'(x){=}u(x)$ for $x\notin X_i$.
\item {\em Synchronization transitions:} If 
$\< l_i,u_i\> \stackrel{!a,g,r}{\mapsto}\< l'_i,u'_i\>$ and  $\< l_j,u_j\>
\stackrel{?a,g,r}{\mapsto}\< l'_j,u'_j\>$ 
then $\tau:=\<\bar{l},u\>\mapsto\<\bar{l}[l_i/l'_i,l_j/l'_j], u'\>$ is a discrete
transition in the network model if $u'(x)=u'_k(x)$ for $k\in\{i,j\}$ and $x\in X_k$, and
$u'(x)=u(x)$ for $x\notin X_k$.
\end{itemize}

We first run the \cipm\ algorithm over each automaton individually. We then compose the
pruned automata to obtain a pruned network. Conjuncting the newly generated
invariants within the individual automata yields new invariants for the network:
\begin{theorem}
\label{theo.parallel}
Assume $\A= \A_1\|\dots\|\A_n$ is a network of timed automata where
$\cipm(\A_i)=(\A'_i,\Ia'_i)$ for each $1\leq i\leq n$, and $\A'=
\A'_1\|\dots\|\A'_n$. Then we will have $\A\eqa\A'$ and $\bigwedge_{1\leq
i\leq n}\Ia'_i(l_i)$ is invariant in $\bar{l}=(l_1,\dots,l_n)$.
\end{theorem}



\begin{example}
\label{example1} 
Figures~\ref{fig.automata1} and~\ref{fig.automata41} show an example of a
timed automaton $\A$ in~\cite{Moller02,Sorea:FTRTFT04}, also the outcome of applying \cipm\ on it.

\begin{figure}
\begin{minipage}[h]{0.4\linewidth}
\center
\input{automata1.pstex_t}
\caption{Example from \cite{Moller02}\label{fig.automata1}}
\end{minipage} 
\begin{minipage}[h]{0.4\linewidth}
\center
\input{automata41.pstex_t}
\caption{After applying \cipm \label{fig.automata41}} \vspace{-2ex}
\end{minipage} 
\end{figure} 
\end{example}

\begin{example}
\label{ex.synch}
The example depicted in Figure~\ref{fig.synch} includes synchronization.
Running the \cipm\ algorithm on $\A_1$ would result in the automaton $\A_2$
depicted in Figure~\ref{fig.synch-end}. The algorithm would not change $\B_1$.
However the parallel composition of $\A_2$ and $\B_1$ would lead to the
parallel automata in Figure~\ref{fig.synch-end}. This is because by
Theorem~\ref{theo.inv} $\A_1\| B_1\eqa A_2\| B_1 $ and according to the
definition of synchronization transitions $A_2\| B_1 \eqa A_2\| B_2$.  As the
figure depicts any configuration of the form $\<(l_i,s_j),u\>$ for $i=4$ or
$j=1$ is unreachable in $\A_2\| \B_2$.  Therefore, according to
Theorem~\ref{theo.parallel} any such configuration is also unreachable in
$\A_1\| \B_1$.

\begin{figure}[htb]
\begin{minipage}[h]{0.4\linewidth}
\center
\resizebox{55mm}{!}{\input{synch.pstex_t}}
\caption{Parallel composition.~\label{fig.synch}} 
\end{minipage} 
\begin{minipage}[h]{0.7\linewidth}
\center
\resizebox{55mm}{!}{\input{synch-end.pstex_t}}
\caption{After applying \cipm .~\label{fig.synch-end}}
\end{minipage} 
\end{figure} 

\end{example}

\section{Predicate Abstraction, New Results and the Ongoing Work}
\label{sec.predicate-abst}

In this section, we introduce a method for using the invariants generated by
 \cipm\ in order to build an over-approximating {\em predicate abstraction} of the 
original timed automaton. 
We consider the abstract states not as Boolean vectors over the designated
set of abstraction predicates, but rather as {\em pairs} of control locations and
conjuncted, positive or negative predicates.  
In the sequel we will explain this in more detail.

A {\em cube} $q$ over $P=\{p_0,...,p_n\}$,
called a {\em minterm} in~\cite{Lahiri06},
is a conjunction 
$\bigwedge_{0\leq i\leq n} \tilde{p_i}$ 
over the elements of $P$ and their negations, 
i.e.\, each $\tilde{p_i}$ is equivalent to either $p_i$ or
its negation $\bar{p_i}$.
For example $x< 0 \land y>2 \land z=3 $ is a cube over
$\{x\geq 0, y\leq 2 , z=3\}$.  $\q(P)$ denotes the set of all cubes over
$P$. In the sequel we assume that $\cipm(\A_1)=(\A,\Ia)$ for a real time model
 $\A_1$, and our intention is to explain how to generate a predicate
 abstraction for $\A_1$. Without loss of generality, in the remainder of the paper 
we use $\Ia(l_i)$ for $\atom(\Ia(l_i))$.

\begin{definition}[States of $\abst_\A$. ]
\label{def.abst}
The set $\I:=\bigcup_{0\leq i< \card{A}} \Ia(l_i)$ is a
collection of all invariants $\Ia(l_i)$. Our predicate abstraction over
$(\A,\Ia)$, denoted $\abst_\A$, is  a finite state automaton where states are
pairs like $(l_i,\bigwedge_{p\in\Ia(l_i)}p\land\bigwedge_{p\in\I\backslash
\Ia(l_i)} \tilde{p})$ for $0\leq i< \card{\A}$. 
\end{definition}

Spurious counterexamples when searching in the abstract state space are often due 
to invariant violations in the concrete model.
In order to reduce the risk of generating spurious counterexamples 
we associate with each control location $l_i$ its invariant
as generated by \cipm . 
These invariants are gathered in $\Ia(l_i)$.
We first pair up each control location to its own invariant. Then we
add the rest of the cubes from $\I \backslash \Ia(l_i)$ to the pair. 
During construction of the abstraction each configuration $\<l_i,u\>$ from the concrete
model is abstracted to a abstract state in which $\Ia(l_i)$ holds.

Let us consider $\q_i$ as the set of all cubes over $\I \backslash \Ia(l_i)$ which are
satisfiable in conjunction with the predicates in $\Ia(l_i)$: 
\[\q_i:=\{q~|~q \in\q(\I \backslash\Ia(l_i))\text{ and }
(\bigwedge_{p\in\Ia(l_i)}p)\land q~\text{is satisfiable}\}. \]

\noindent
For each $q\in \q_i$ we denote by $[l_i, q]$ the abstract state 
$(l_i,(\bigwedge_{p\in\Ia(l_i)}p)\land q )$.
$[l_i, q]$ abstracts all configurations $\< l_i,u_i\>$ in the concrete
model $\A$ whose valuation $u_i$ satisfies $q$, i.e.\ $u_i\models
 q$.

\begin{example}
Let us continue with the first example (Figure~\ref{fig.automata41}).
According to the example, we have $\Ia(l_0)=\{y\leq 1\}$,
$\Ia(l_1)=\{x\leq y\}$, $\Ia(l_2)=\{y<x\}$
 and hence, $\I=\bigcup_{0\leq i<
\card{A}} \Ia(l_i)= \{y\leq 1, x\leq y, y<x\}$. 
We use $p_i$ to denote the invariant corresponding to the location $l_i$, therefore:
$\q(\I \backslash\Ia(l_0)) =\{p_1\land p_2, \bar{p_1}\land p_2 , p_1\land
\bar{p_2}, \bar{p_1}\land \bar{p_2}\}  $
$\q(\I \backslash\Ia(l_1)) 
=\{p_0\land p_2, \bar{p_0}\land p_2 , p_0\land \bar{p_2}, \bar{p_0}\land
\bar{p_2}\} $
$ \q(\I \backslash\Ia(l_2))  =\{p_0\land p_1, \bar{p_0}\land p_1
, p_0\land \bar{p_1}, \bar{p_0}\land \bar{p_1}\}$
Some of these combinations are unsatisfiable, for instance $p_1\land p_2$.  
After removing such combinations and eliminating the
'$\land$' symbol, for simplicity, we obtain: $\q_0 =\{\bar{p_1} p_2 , p_1
\bar{p_2}\}$, $\q_1 =\{p_0 \bar{p_2}, \bar{p_0} \bar{p_2}\}$, and $\q_2 =\{p_0
\bar{p_1}, \bar{p_0} \bar{p_1}\}$.
As illustrated in Figure~\ref{fig.abst1} these three sets build an abstract
model $\abst_A$ which consists of six states for example like $(l_0,p_0p_1\bar{p_2})$, 
$(l_1,p_1p_0\bar{p_2})$. As we shall see later
on, the dashed line in this figure identifies unreachable states.
\end{example}

\begin{definition}[Transitions of $\abst_\A$]{\em
In $\abst_\A$ we execute a transition from a state $[l_i,q]$ to a state
$[l_j,q']$ only when one of the following conditions holds in the concrete
model $\A$:
\begin{itemize}
\item there are two valuations $u_i$ and $u_j$ and a non-idle transition $\<
l_i,u_i\> \stackrel{\tau}{\mapsto}\< l_j,u_j\>$ where $u_i\models q$ and
$u_j\models q'$, or
\item $l_j$ is identical to $l_i$, and there is a delay transition $\<
l_i,u_i\> \stackrel{d}{\mapsto}\< l_i,u_i+d\> $ for some valuation $u_i$ such
that $u_i\models q$ and $u_i+d\models q'$.
\end{itemize}}
\end{definition}

\noindent
Let $\next([l_i, q])$ denote the set of all successor states of $[l_i,q]$ in
$\abst_\A$, then with respect to 
definition above:
\begin{align*}
\next([l_i, q]):= & \{[l_j, q']~|~\exists\tau ~\text{or}~d:~
                  \< l_i,u_i\> \stackrel{\tau / d} {\mapsto}\< l_j,u_j\>
                  \text{ such that } \\
                  & u_i\models (\bigwedge_{p\in\Ia(l_i)}p)\land
                  q \text{ ~and~ } u_j\models (\bigwedge_{p\in\Ia(l_j)}p)
                  \land q'\}. ~\hspace{1cm} (2)
\end{align*}
Recall that $\tau$ is a discrete and $d$ is a delay transition.

Since $\abst_\A$ is an abstraction of $\A$, each of its transitions should 
have a counterpart in the original model $\A$. This means that whenever 
$[l_j, q']\in \next([l_i, q])$, there must exist a non-idle transition from
at least one of the corresponding concrete states of $[l_j, q]$ to that of $[l_j, q]')$. 
Such a transition needs to satisfy all the invariants of the source location and also
all the invariants of the target location. Also if there is a reset for some variable, the new
value of the respective variable should satisfy the invariant of the target location:

\begin{lemma}
\label{lem.trans}
Assume that $\abst_\A$ is an abstraction of $\A$ with respect to some set of
predicates $P$.  There is a transition from $[l_i, q]$ to $[l_j, q']$ in
$\abst_\A$, i.e. $[l_j, q']\in \next([l_i, q])$, if and only if 
one of the conditions below holds:
\begin{enumerate}
\item there are two clock valuations $u_i$ and $u_j$, and a non-idle transition 
$\tau:\< l_i,u_i\> \mapsto\< l_j,u_j\>$ in the concrete model such that:
  \begin{enumerate}
   \item $u_i\models q$ and $u_j\models q'$.
   \item if $\G_{\tau}\neq \emptyset$ then $\G_{\tau} \land q$ is satisfiable, 
   \item if $\gr\neq \emptyset$ then $\gr \land q'$ is satisfiable, 
   \item if $\R_{\tau}\neq \emptyset$ then $\overline{\atom}(\R_{\tau}) \land q'$ is satisfiable, 
   \item for all variables $x\notin var(\R_{\tau})\cup var(\G_{\tau})$, $u_i(x)=u_j(x)$.
  \end{enumerate}
\item $l_i=l_j$ and $\exists d,u_i:~\< l_i,u_i\>{\mapsto}\<l_i,u_i+d\>$ where
  $u_i\models q$ and $u_i+d\models q'$.
\end{enumerate}
\end{lemma} 
The next theorem shows that in order to establish a predicate abstraction for the
original concrete model $\A_1$ it is enough to do so for the pruned
equivalent version obtained from an application of the $\cipm$ algorithm:

\begin{theorem}
\label{theo.approx}
If $\cipm(\A_1)=(\A,\Ia)$, then $\abst_\A\eqa \abst_{\A _1}$. \vspace{-0ex}
\end{theorem}

\noindent
The cube $p_0 p_1 \bar{p_2}$ has caused two
different abstract states in Figure~\ref{fig.abst1}. This is because $p_0$ and $p_1$ are invariants of 
$l_0$ and $l_1$, respectly, and therefore coupled with them in the abstract model.
The dashed line in this figure depicts the set of unreachable
abstract states of the first example. These states are unreachable since they correspond to some 
unreachable concrete states in $\A$ (cf. Lemma~\ref{lem.trans}).
Using Lemma~\ref{lem.trans} to compute the transitions in the abstract model, 
one would obtain Figure~\ref{fig.abst3} as the initial predicate abstraction
of $\A$. For instance from $(l_0,p_0\bar{p_1}p_2)$ there is a transition to
$(l_0,p_0p_1\bar{p_2})$ because the transition $\<l_0,u\>
\stackrel{x:=0}{\mapsto}\< l_0,u'\>$ fullfils  Lemma~\ref{lem.trans}.

\begin{figure}
\begin{minipage}[h]{0.5\linewidth}
\center
\input{abst1-new.pstex_t}
\caption{The states of $\abst_\A$~\label{fig.abst1}} \vspace{-2ex}
\end{minipage} 
\begin{minipage}[h]{0.5\linewidth}
\center
\input{abst3.pstex_t}
\caption{~\label{fig.abst3}$\abst_\A$, predicate abstraction of $\A$. } 
\end{minipage} 

\end{figure}

In the following we give a simple succinctness analysis of our approach: 
Each timed automaton has a finite number of control locations, $\card{\A}$.
We associate with each location $l_i$ at most $\card{\q_i}$ abstract
states. This way the number of the abstract states is at most $\Sigma_{0\leq
i<\card{\A}} \card{\q_i}$ in the worst case. In the example depicted in
Figure~\ref{fig.abst1}, this number is $2+2+2=6$.
By pruning the original model using \cipm\ and also with respect to
Lemma~\ref{lem.trans} this number reduces to $4$ abstract states, see Figure~\ref{fig.abst3}.
With neither detecting the idle transitions nor pairing the control locations
with their invariants, in the abstraction facet, one would have gotten
$3\times 4=12$ abstract states where $4$ is the number of distinguished
satisfiable cubes and $3$ is
the number of control locations. This number would have even raised to $3\times
2^3=24$ abstract states if no satisfiability check on the cubes was done. 
\vspace{-3.8ex}

\bibliographystyle{plain}
\bibliography{bib}

\end{document}